\documentclass[preprint,a4paper]{revtex4-1}

\usepackage{hyperref} 
\usepackage{amsmath}
\usepackage{graphicx}
\usepackage[utf8]{inputenc}
\usepackage{color}
\usepackage{caption}
\usepackage{subcaption}
\usepackage{natbib}

\DeclareMathOperator{\e}{e}
\newcommand{\ud}{\,\mathrm{d}}

\begin{abstract}
Lipid monolayers and bilayers have been used as experimental models
for the investigation of membrane thermal transitions. The main transition
takes place near ambient temperatures for several lipids and reflects
the order-disorder transition of lipid hydrocarbonic chains, which is accompanied
by a small density gap. Equivalence between the transitions in the two systems
has been argued by several authors. 
The two-state statistical model adopted by numerous authors for different properties
of the membrane, such as permeability, diffusion, mixture or insertion of cholesterol
or protein, is inadequate for the description of charged membranes, since it lacks a 
proper description of surface density. We propose a lattice solution model which
adds interactions with water molecules to lipid-lipid interactions
and obtain its thermal properties under a mean-field approach. Density variations,
although concomitant with chain order variations, are independent of the latter.
The model presents both chain order and gas-liquid transitions, and extends the
range of applicability of previous models, yielding Langmuir isotherms in the full
range of pressures and areas.
\end{abstract}

\begin{document}

\title{Lattice solution model for order-disorder transitions in membranes and Langmuir monolayers}
\date{\today}
\author{Henrique S. Guidi}
\email[E-mail me at: ]{henrique@if.usp.br}
\affiliation{Instituto de Física da Universidade de São Paulo}

\author{Vera B. Henriques}
\email[E-mail me at: ]{vera@if.usp.br}
\affiliation{Instituto de Física da Universidade de São Paulo}

\maketitle

\section{\label{sec:intro}Introduction}

Lipid monolayers \cite{kaganer1999, mohwald1995} and bilayers \cite{bloom1991, nagle2004} 
have been extensively used as experimental models for
the investigation of thermal and structural properties of the biological membrane. 
Phospholipid molecules form an ordered monolayer film
on the air-water interface, with lipid headgroups resting on water, while lipid hydrophobic chains
acquire approximately parallel orientation in the air phase, thus avoiding contact with
water. External lateral pressure guarantees
aggregation into the monolayer film.
In water solution,
lipids aggregate into bilayer vesicles, as polar heads shield hydrocarbonic tails from contact with
the aqueous medium. Bilayers are tension-free and aggregation is driven by the hydrophobic effect.

Both systems may undergo several phase transitions.
One of the most thoroughly investigated is the pronounced 
order-disorder lipid chain transition,
with latent heat,
presented by either system.
For lipid membranes
temperature or pH variations may yield a so called main gel-fluid transition.
In the case of lipid monolayers,
compression or heating disclose a transition traditionally known as a liquid-condensed 
liquid-expanded transition.
An abrupt variation in lipid surface density is accompanied
by disordering of the hydrocarbon chains. 
The latter acquire ``kinks'',
while distance between polar headgroups increases,
yielding decreased surface density.
This is recognized as the chief effect also for bilayers \cite{nagle1973, nagle1975, marcelja1974},
which undergo the main transition under temperature-variation.

There are different advantages in
adopting one experimental model or the other. Direct measurement of surface area per lipid molecule,
whose abrupt variation signals the transition, is possible only for monolayers. However,
the discontinuity in lipid area depends on external applied pressure.
Equivalence between the two systems for a particular pressure on monolayers has been
argued by many authors \cite{nagle1976, gruen1982, marsh1996} and arguments rest on the assumption of
negligible interaction between the two leaflets that compose the bilayer.
However,
recent studies indicate that this might be too strong a hypothesis \cite{nagle2004}.
Thus differences should be subject of further investigation.

Lipids on the air-water interface reduce the surface tension,
as lipid headgroups disrupt 
the specially stable hydrogen bonds between surface water molecules \cite{ni2013}. 
Bilayers are free of internal surface pressure. The hydrophobic effect, which consists of
'micro' phase-separation between lipids and water, as a consequence of the disparity
between the values of Van der Waals attraction between aliphatic chains as compared to
the value of water hydrogen bond energies, 
 is considered to be the main force driving
membrane behaviour.  This effect is absent in the case of monolayers, since 
 chains avoid water by turning to the air subphase. 

In this study,
we approach some of the questions related to the main transition
from the point of view of a minimal statistical model.
Different two-state and multi-state models,
inspired on the success of Ising-like model for magnetism, 
were proposed by several 
authors in the 1970's-1980's \cite{caille1978, doniach1978, mouritsen1983}, reflecting the possible states of the lipid chains,
either extended (all-trans) or disordered (gauche kinks) by different degrees, 
with focus on the different areas occupied by the lipid head-groups on the bilayer surface. 
The orientational order of lipids in the layer also suggested inspiration
on models for nematic liquid crystals. On the other hand, the large enthalpy attributed to
chain melting suggested that this would be the main entropic mechanism for the transition,
thus demanding accurate treatment of chain configurations, an approach followed by Nagle
\cite{nagle1973}.

Marcelja \cite{marcelja1974} proposed that
chain kinks could be treated in terms of a nematic-like order parameter along
the chain, subject to an effective field due to density of extended chains. Thus
chain entropy was obtained from the statistical calculation. Area per lipid headgroup
was taken as linearly dependent on the inverse of lipid chain length, from molecular
volume conservation. 
Caille \cite{caille1978} considered a lattice of two-state particles,  corresponding to
the ordered and disordered chains. The disordered chain lipid would occupy a 
certain number of sites, and was attributed an intramolecular chain
entropy.
Some authors looked also at multi-state models \cite{caille1978, mouritsen1983}.  
A lattice-gas three-state model, with two states for the lipids,
 was also suggested \cite{baret1983}, 
but the authors left out the essential intra-chain degeneracy.
Nagle \cite{nagle1973} proposed a different approach:
chain configurations of a two-dimensional section of the lipid monolayer could be exactly enumerated
through mapping on a dimer counting problem.
However,
the much simpler treatment of chain entropy in terms of an average
degeneracy of the two-states model came to dominate the literature.

Doniach \cite{doniach1978} simplified  the two-state
model of \cite{caille1978} by associating to each
a different area
parameter in an {\it ad hoc} fashion,
and noticed the possibility of treating the
resulting model
exactly, through mapping on the seminal 
two-dimensional Ising model.
Doniach's version of the original two-state model 
turned into the most successful statistical model 
for the main transition of the
 lipid system.
Inspite of its success for
the description of the transition for multicomponent lipid membranes, for diffusive properties,
or for the effect of protein or cholesterol insertion 
\cite{mouritsen1983, pink1979, bloom1991, heimburg, almeida2011}, 
the model focuses on the order-disorder transition of the lipid chains,
while  the area per chain is introduced in an {\it ad hoc} manner,
by attributing different areas to the 'ordered' and 'disordered' sites.
As a consequence, a true discontinuity in density is not displayed
by the model at the order-disorder transition of the chains. This
aspect represents an important limitation
 in the case of competing interactions, 
such as for ionic lipid layers \cite{tamashiro2011, LamyFreund2003, Barroso2010}.
For dissociating lipids, electrostatic repulsion competes with the hydrophobic effect,
and a delicate equilibrium is established between chain order, charge and molecular 
surface density. In this case, precise description of the local lipid  density
 is essential in order to appropriately rationalize thermal, electrical and structural
properties of the experimental system \cite{henriques2011, tamashiro2011}.

In this study we have considered a two-state lipid lattice solution in which
sites may be occupied either by lipid or by water particles. 
Chains may be in two different states, one of them largely degenerate,
but the density results from the equilibrium occupation of the lattice.
Thus area per lipid is obtained from the statistics of the model. Also,
proper treatment of pressure allows examination of the equivalence hypothesis of mono
and bilayers.

While recognizing
the fact that different authors contributed to the formulation of the
original two-state model, in the name of simplicity
we shall identify our model as a lattice solution Doniach model.
In section 2, we define the statistical model. In section 3, we present our mean-field 
approach.  Results for thermodynamic properties
and phase diagrams are displayed in section 4 . 
Physical interpretation in terms of the two systems of interest,
monolayers and bilayers, is discussed in section 5. Final comments 
are in section 6.


\section{Definition of the statistical model}

\begin{figure}
\includegraphics[scale=0.9]{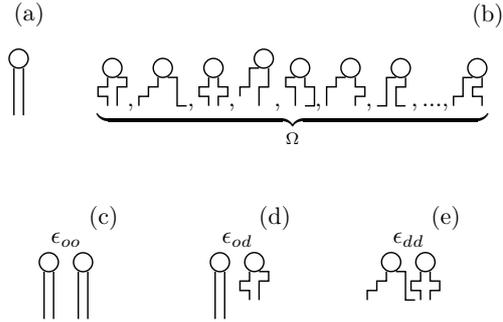}
\caption{(a) Simplified representation of a lipid with extended hydrocarbon chains.
(b)  Disordered chain configurations are represented through a single average disordered state
 of  degeneracy $\Omega$.
(c)-(e) interaction between pairs of lipids in different states.}
\label{fig:termos}
\end{figure}

We revise the seminal model proposed by Doniach a few decades ago 
\cite{doniach1978, mouritsen1983, bloom1991}
for the phospholipid bilayer main transition in order to set notation.
In Doniach's lattice model,
lipid chains fill the plane lattice and are considered to visit two different particle states,
an {\it{ordered chain}} state $o$ (Fig. \ref{fig:termos}a),
corresponding to an extended chain,
and a highly degenerate  {\it{disordered chain}} state $d$,
meant to represent an average shortened chain (Fig. \ref{fig:termos}b).

The system consists of $N$ particles distributed over the
 $L^2=N_\text{o} - N_\text{d}$ sites.
Its configurational energy may be written as
\begin{equation}
 E =
-\epsilon_\text{oo} N_\text{oo}
-\epsilon_\text{dd} N_\text{dd}  
-\epsilon_\text{od} N_\text{od}
\label{eq:Doniach_energy}
\;\text,
\end{equation}
where $N_{xy}$ is the number of contacts between two particles in states $x$ and $y$,
 and $N_{x}$ is the number of lipids in state $x$,
where $x=\{\text{o},\text{d}\}$  and $y=\{\text{o},\text{d}\}$.
Interaction parameters $\epsilon_{xy}$ should all be taken as positive,
since they represent effective attraction between particles.
At the main transition,
there is a sharp variation of the lipid chain states.
A chain order parameter 
\begin{equation}
m=\frac{N_\text{o} - N_\text{d}}{L^2}
\label{eq:chain_parameter}
\end{equation}
describes chain order.

The model incorporated a second feature.
Disordering of the chains is intimately related physically to loosening of the packing of lipid headgroups,
as indicated by experiments.
It seemed natural to make lipid area dependent on lipid chain state.
Thus in Doniach's proposal a lipid particle in the ordered state could be associated with surface particle area $a_\text{o}$,
while a lipid molecule in one of the disordered states would be given area $a_\text{d}$,
with $a_\text{d} > a_\text{o}$. As a consequence, lattice and model areas have
no correspondence, with
$A \ge L^2$, while $N_{lip}=L^2$.
This approach implies that the area per particle $a_{\text{Doniach}}$ is defined as
%
\begin{equation}
a_{\text{Doniach}} =
\frac{N_\text{o} a_\text{o} + N_\text{d} a_\text{d}}{L^2}
\;\text.
\end{equation}
%
As it can be seen, 
chain order parameter $m$ and area per particle $a_{\text{Doniach}}$ are not independent thermodynamic variables,
since
%
\begin{equation}
 a_{\text{Doniach}}  ( m ) =
\frac{1}{2}  m  (a_\text{o} -a_\text{d})
+ \frac{1}{2} (a_\text{o} + a_\text{d})
\;\text.
\label{eq:a(m)}
\end{equation}
%

The model simplicity allows mapping on a modified form of the two-state Ising magnetic model,
whose thermodynamic properties are very well established.
Lateral pressure,
conjugate to 'area', acts as en effective field favouring the ordered phases. 
Together with temperature, this field 
controls the lipid system phases.
Chains are ordered at large lateral pressures and low temperatures,
as expected.
At fixed temperatures,
chains order discontinuously under increasing lateral pressure.
As temperature increases,
the transition disappears at a critical temperature.

Despite the utility of the model for many purposes,
it may be unsuitable under certain circumstances,
such as in the case of charged lipid membranes \cite{tamashiro2011},
whose thermodynamic properties depend strongly on charge surface density.

\subsection*{Doniach's lattice solution model - DLG}

We propose to introduce lipid density as a true statistical variable and write Doniach's model 
as a lattice solution  with explicit water particles.
The inclusion of interactions between lipid and water particles is essential for
the purpose of investigation of the bilayer-monolayer analogy 
\cite{nagle1976} and Marsh\cite{marsh1996}.
Like with its predecessor,
our goal is to describe the order-disorder transition of
a surface of lipid chains within the framework of a simplified model.

\begin{figure}
\centering
\includegraphics[scale=0.9]{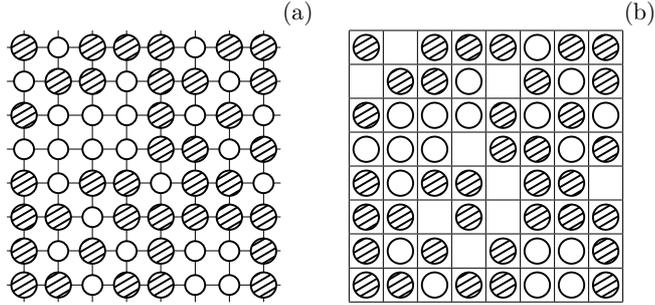}
\caption{(a) Illustration of Doniach's model.
Lipid states define area per lipid, which is independent of lattice spacings. 
(b) A lattice gas version of Doniach's model (DLG).
Sites are occupied either by lipid or water particles.
Area per lipid is obtained from model statistics.
}
\label{fig:2}
\end{figure}

In this new proposal,
the fixed relation between area per particle and chain parameter $m$, Eq. \ref{eq:chain_parameter},
is abandoned.
Figure \ref{fig:2} illustrates pictorially our proposal,
as compared to the original Doniach description.

Let us consider a square lattice of area $A$ and $L^2=A$ sites,
which may be 
occupied by lipids or by water.
We define occupational variables $(\delta=1)$ and $(\delta=0)$,
for lipids and water,
respectively.
The lipid particle chains may be either in the ordered or disordered state,
with chain variables given by $(\eta=1)$ or $(\eta=0)$,
accordingly.
Interactions between lipid particles are the same as those of Eq. \ref{eq:Doniach_energy},
but the number of lipid particles $N_\text{lip}=N_\text{o} + N_\text{d}$
is not fixed.  Model energy reads
\begin{eqnarray}
 E & = &
-\epsilon_\text{oo} N_\text{oo}
-\epsilon_\text{od} N_\text{od}
-\epsilon_\text{dd} N_\text{dd}  \nonumber \\
& &
-\epsilon_\text{ow} N_\text{ow}
-\epsilon_\text{dw} N_\text{dw}  
-\epsilon_\text{ww} N_\text{ww}  
\label{eq:DLG_energy}
\;\text,
\end{eqnarray}
where $N_\text{xy}$ is the number of contacts between sites
occupied by lipids in ordered state $o$,
by lipids in disordered state $d$
or by water particles $w$.
As in Doniach's model,
the disordered states are multiply degenerate,
the degeneracy 
$\Omega$ 
corresponding to the high entropy of
the disordered hydrocarbon chains of a single lipid.

In order to give our model a statistical treatment, it is convenient to 
rewrite energy (Eq. \ref{eq:DLG_energy}) in terms of  statistical variables. We attribute
variables $\sigma$ to lattice sites as in table \ref{tab:one}. Under this notation, Eq. \ref{eq:DLG_energy}
is rewritten as
%
\begin{equation}
E  =  
\sum_x
\sum_y
-\epsilon_{xy} 
\sum_{(ij)} \eta_i^x \eta_j^y
\end{equation}
%
where $\eta$ is defined in table \ref{tab:one}.

\begin{table}
\caption{\label{tab:one}Particle states and statistical variables}
\begin{ruledtabular}
\begin{tabular}{ccc}
particle & occupation variable $\sigma$ & mapping $\sigma \to \eta$\\
ordered chain lipid (o)& $1$ & $\frac12 \sigma^2(1+\sigma)$\\
disordered chain lipid (d) & $-1$ & $\frac12 \sigma^2(1-\sigma)$\\
water (w)& $0$ & $1-\sigma^2$\\
\end{tabular}
\end{ruledtabular}
\end{table}

Under this representation, and
after some manipulation,
system energy (Eq. \ref{eq:DLG_energy}) is given by
an interaction term $E_\text{int}$ and
a hydrophobic "field" term $E_\text{hydr}$, besides a constant term:
\begin{eqnarray}
E&=&  E_\text{int} + E_\text{hydr} + E_{0} \nonumber \\
& = &  
-J
\sum_{(ij)} 
\sigma_i \sigma_j
-\Delta
\sum_{(ij)} 
\sigma_i \sigma_j \left(\sigma_i +\sigma_j\right)
\nonumber \\
&&
-K
\sum_{(ij)} 
\sigma _i^2 \sigma_j^2
+ I
\sum_i \sigma_i^2
- 2 \epsilon_\text{ww} A
\;\text,
\label{eq:vh:hamiltoniana}
\end{eqnarray}
%
where,
for simplicity,
new interaction parameters are defined as
\begin{equation}
J = 
\frac{\epsilon_\text{oo} + \epsilon_\text{dd} - 2 \epsilon_\text{od}}{4} 
\;\text,
\label{eq:parameters:J}
\end{equation}
\begin{equation}
\Delta=
\frac{\epsilon_\text{oo} -\epsilon_\text{dd} - 2\epsilon_\text{ow} +2\epsilon_\text{dw}}{4} 
\;\text,
\label{eq:parameters:Delta}
\end{equation}
\begin{equation}
K = 
\frac{\epsilon_\text{oo} + \epsilon_\text{dd} + 2 \epsilon_\text{od} + 4\epsilon_\text{ww} -4 \epsilon_\text{ow} -4\epsilon_\text{dw}}{4} 
\label{eq:parameters:K}
\end{equation}
and
\begin{equation}
I  = 
 2( 2 \epsilon_\text{ww} - \epsilon_\text{ow} - \epsilon_\text{dw})
\;\text.
\label{eq:parameters:I} 
\end{equation}

For the lattice solution lipid model,
equilibrium properties are more easily calculated in the grand-canonical ensemble.
The grand-partition function reads
\begin{equation}
\Xi(T,\mu_\text{lip}, \mu_\text{w})=\sum_{\{\sigma\}}
\Omega ^ {N_\text{d}}
\e^{-\beta \left( E - \mu_\text{lip} N_\text{lip} - \mu_\text{w} N_\text{w} \right) }
\label{eq:part_function}
\;\text,
\end{equation}
where $\mu_\text{lip}$ and $\mu_\text{w}$ are the lipid and water chemical potentials.
The total number of lipid particles $N_\text{lip}$, the number of water particles
$N_\text{w}$ and the number of disordered chain lipids $N_\text{d}$ are
given respectively by
\begin{equation}
N_{\text{lip}} = \sum_{i}\sigma_{i}^2
\label{eq:N}
\end{equation}
and
\begin{equation}
N_\text{w}= A - N_\text{lip}
\label{eq:Nw}
\end{equation}
and
\begin{equation}
N_d = \sum_{i}\sigma_{i}^2 (1-\sigma_i)/2
\;\text.
\label{eq:Ndis}
\end{equation}

It is interesting,
at this point,
to note that the linear "hydrophobic" energy term
in the energy expression (Eq. \ref{eq:vh:hamiltoniana})
competes with the chemical potential factor, so we rewrite
the grand-partition function as 
\begin{equation}
\Xi(T,\mu)=\e^{\beta (\mu_\text{w}+2\epsilon_\text{ww}) A}
\sum_{\{\sigma\}}
\Omega^{N_\text{d}}({\sigma})
\e^{-\beta 
\left( 
E_\text{int}(\{\sigma\}) - \mu N_\text{lip}(\{\sigma\}
\right)}
\label{eq:part_function}
\;\text,
\end{equation}
where  $\mu \equiv  \mu_{\text{lip}} - \mu_\text{w} - I$.

\section{Mean-field approach}

The model equilibrium properties may be obtained from a 
Curie-Weiss mean-field approach
\cite{carneiro1989}
which allows linearization of the system energy in the
 statistical sums and turns exact calculation possible.
Interactions are made independent of distance,
with the replacement
\begin{equation}
\sum_{(i,j)} X_i X_j \to \frac{q}{2 A}\sum_i X_i \sum_j X_j
\;\text,
\end{equation}
where $X_i$ are interaction variables,
$A=L^2$ is the system area and $q$ is the model coordination number.
Under this transformation,
the model energy is written as:
\begin{equation}
E_\text{MF} = 
\frac{q}{2 A}
\left(
-J M^2
-2\Delta MN
-K N^2
+ I N \frac{A}{2}
- \epsilon_{\text{ww}} A^2
\right)
\;\text,
\label{eq:mf_energy}
\end{equation}
where $N$ is the global number of particles given by Eq. \ref{eq:N},
and $M$ is defined as
\begin{equation}
M = \sum_{i} \sigma_i
\;\text.
\end{equation}

Linearization of the quadratic terms in $M$ and $N$ in the grand-partition 
function of Eq. \ref{eq:part_function},
with model energy $E$, Eq. \ref{eq:vh:hamiltoniana},
replaced by mean-field energy,
$E_\text{MF}$ (Eq. \ref{eq:mf_energy}),
may be achieved through Gaussian transformations
\begin{equation}
e^{y^2}= \frac{1}{\sqrt{\pi}}
\int_{-\infty}^{\infty}
\e^{-x^2 + 2 y x}
\ud x
\;\text.
\end{equation}

Summation over statistical variables $\sigma$ becomes straightforward.
Integrals in the newly introduced variables $X$ are then solved by the steepest descent method.
For large systems the main contribution comes from extrema which 
yield the following grand-potential $\Phi(T,A,\mu,H)$:
\begin{eqnarray}
\frac{\Phi(T,A,\mu,H;m,n)}{A} & = & 2( J m^2 + K n^2 + 2 \Delta n m ) \nonumber\\
& & 
- 2\epsilon_{\text{ww} - \mu_{w}}
 - \frac1\beta \ln\left(
1+ \phi_+
\right)
\;\text.
\label{eq:cm_f}
\end{eqnarray}

Here $m,n$ are respectively chain order parameter
$m=\left< M \right>/{A}$ 
and lipid densities 
$n=\left< N_{lip} \right>/{A}$.  
$\phi_+$ is a function of $m$ and $n$, defined as
\begin{eqnarray}
\begin{bmatrix}
\phi_+\\
\phi_-
\end{bmatrix}
&=&
\e^{\beta (4 \Delta m + 4 K n + \mu - \frac12 I)} \times 
\label{eq:cm_phi}
\\
&& \left(
\e^{\beta ( 4 J m + 4 \Delta n )}
\begin{bmatrix}
+\\-
\end{bmatrix}
\Omega \e^{-\beta (4 J m + 4 \Delta n )}
\right)
\nonumber 
\;\text.
\end{eqnarray}

The chain order parameter $m(T,\mu)$ and model density $n(T,\mu)$,
are then given through the 
following system of coupled equations which define the conditions for the extrema for exponents
of the Gaussian integrals:
\begin{equation}
\begin{bmatrix}
m\\
n
\end{bmatrix}
(T,\mu;m,n)
=
\begin{bmatrix}
\phi_-
\\
\phi_+
\end{bmatrix}
\frac{1}{1+\phi_+}
\;\text.
\label{eq:cm_mn}
\end{equation}

\section{Model thermodynamic properties and phase diagram}

\begin{figure}
\centering
\includegraphics[scale=0.9]{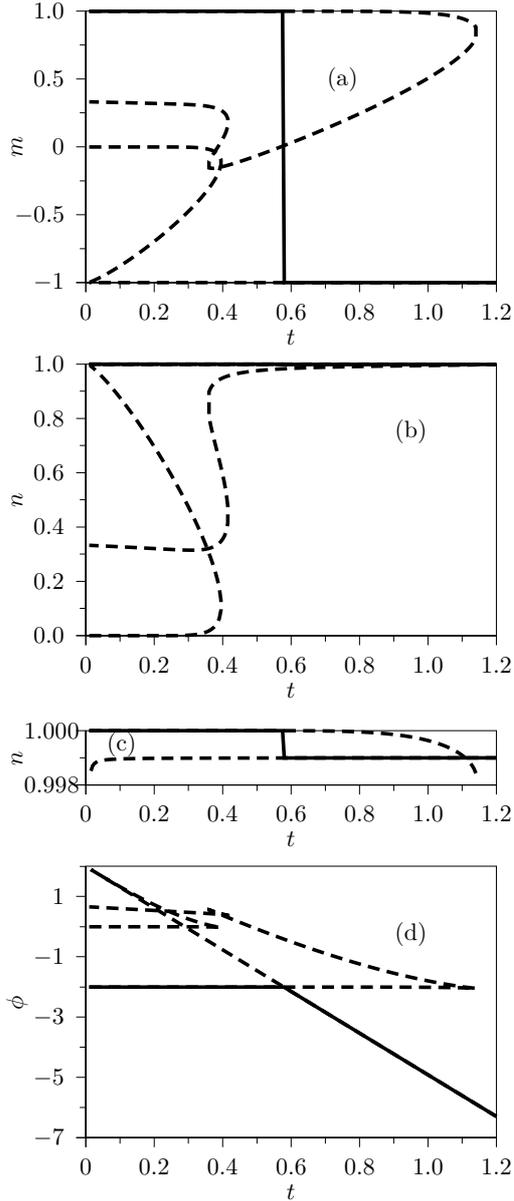}
\caption{Thermal behaviour of chain order parameter $m$ (a) and of density $n$ (b).
A detail of the discontinuity in density is displayed in (c),
and behaviour of free-energy as a function of temperature in (d).
Dashed lines represent all the possible solutions for Eqs.
\ref{eq:cm_mn} and \ref{eq:cm_phi}.
Continuous lines correspond to the absolute minimum of the free-energy.}
\label{fig:fig04}
\end{figure}

Model properties are investigated through inspection of the solutions for
chain order parameter $m$ and lipid density $n$ 
(Eqs. \ref{eq:cm_phi} and \ref{eq:cm_mn}).
Depending on the thermodynamic parameters,
several solutions may be found,
and the equilibrium physical solution is obtained 
from inspection of the global minimum of the grand-potential $\Phi$.
Fig. \ref{fig:fig04}a-d illustrates this procedure. 
Differently from the original Doniach model,
it can be seen that the disordering transition of the chains,
signalled by the abrupt discontinuity in chain parameter $m$,
is accompanied by a small discontinuous transition in density $n$,
shown in the detail. 

\begin{figure}
\includegraphics[scale=0.9]{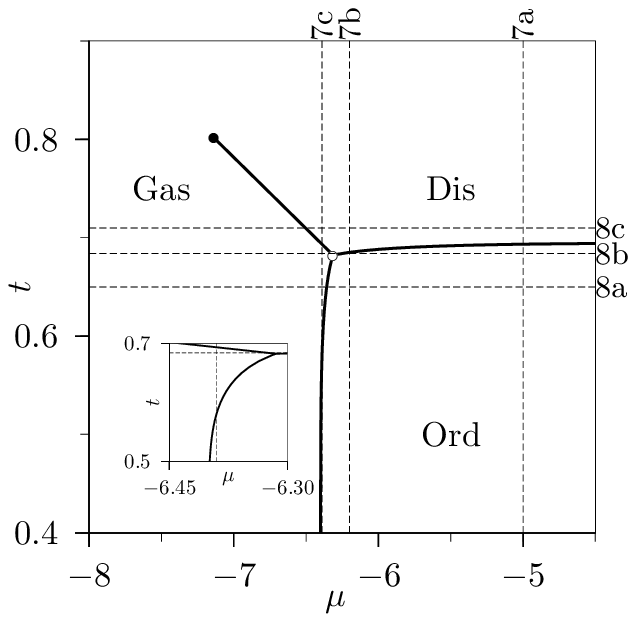}
\caption{
Model phase diagram in the temperature $t/J$ \it{vs} chemical potential $\mu/J$ 
plane. Model parameters are $\Delta/J=0.6$, $K/J=1$ and  $\Omega=1000$.
Continuous lines are coexistence lines. Hollow circle is a triple point.
The gas-disordered chains coexistence line ends at a critical point indicated by a full circle.
Dashed lines represent cuts illustrated in Figs. \ref{fig:fig06} and \ref{fig:fig07}.
}
\label{fig:dia_primeiro}
\end{figure}

A possible phase diagram is displayed in Fig. \ref{fig:dia_primeiro} 
for a specific set of parameters.
Three phases are present: a gas phase (Gas),
characterized by very low density ($n\approx0$);
a liquid of ordered chains (Ord),
with chain parameter $m\approx1$;
and a liquid of disordered chains (Dis), with chain parameter $m\approx-1$.
The gas phase is present at low chemical potentials. 
At low fixed temperature,
a Gas-Ord transition takes place as chemical potential is increased.
For high chemical potentials,
as one increases temperature,
a discontinuous Ord-Dis occurs,
with a small density gap in density $n$ accompanying a sharp transition in chain order parameter $m$,
from $1$ to $-1$.
For a range of intermediate chemical potentials,  
raising temperature produces a Gas-Dis discontinuous transition in density $n$.
The three phases coexist at a triple point.

Reentrant behaviour is displayed by the model system,
near the triple point,
in a small range of chemical potentials.
This is shown in the detail of Fig. \ref{fig:dia_primeiro}:
as temperature is increased,
the ordered chains give place to a gas phase,
and as temperature is increased further,
the gas phase presents coexistence with a disordered chains liquid.

\begin{figure}
\centering
\includegraphics[scale=0.85]{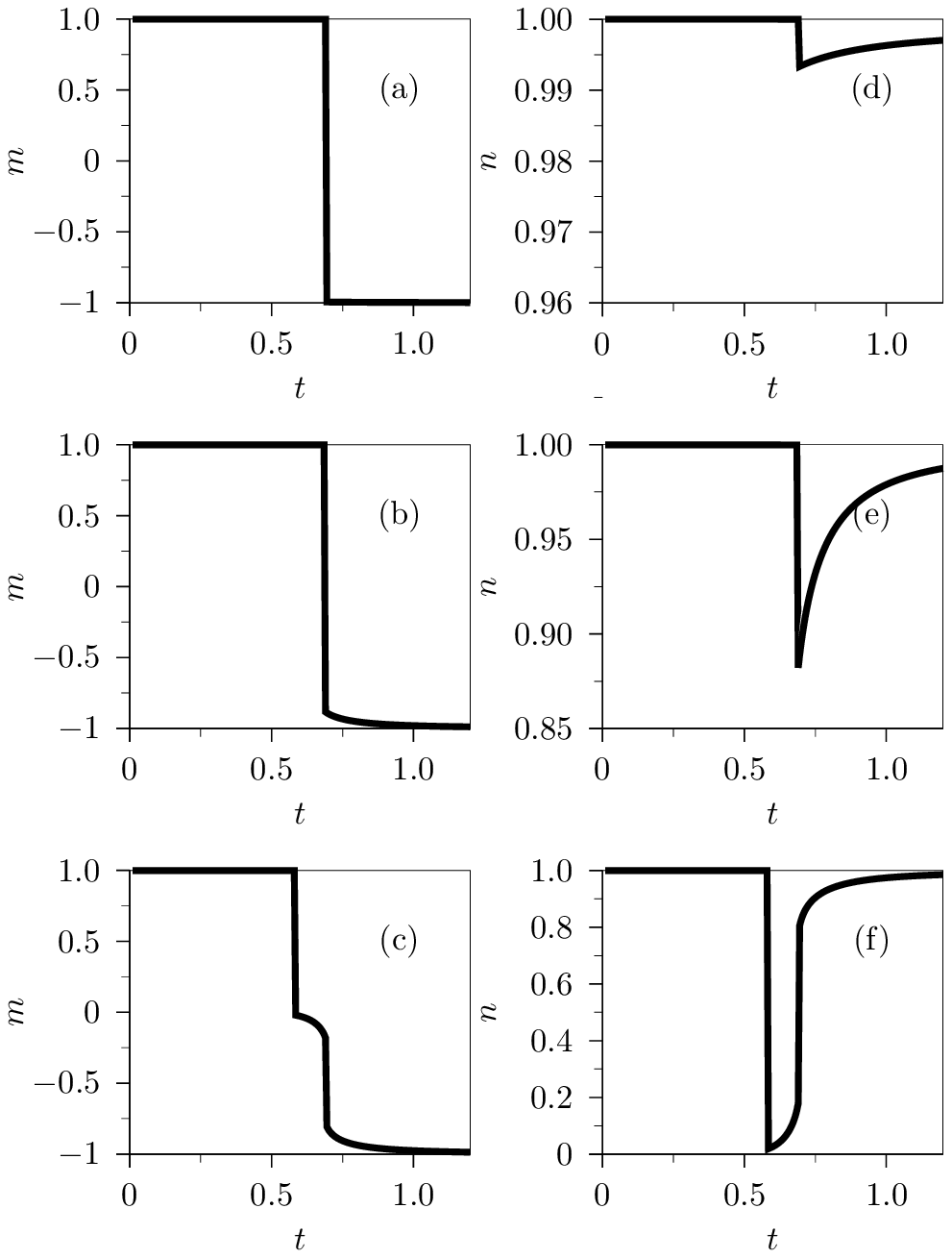} 
\caption{Chain order parameter $m$ (a-c) and 
lipid density $n$ (d-e) as functions of temperature $t$, 
for fixed chemical potentials $\mu$ $-5$, $-6.2$ and $-6.39$.
Figs (a) and (d) illustrate that at higher chemical potentials 
the discontinuous chain parameter transition 
is accompanied by a small density discontinuity. Figs (b) and (d) display model behavior
nearer to the triple point, with a larger discontinuity of density at the order-disorder transition.
Figs (c) and (e) show reentrant behaviour beyond the triple point, with  
an order-gas transition and a gas-disordered chains transition as temperature is raised. 
The three cuts above are shown as 
dashed lines in the phase diagram of Fig. \ref{fig:dia_primeiro}.} 
\label{fig:fig06}
\end{figure}

\begin{figure}
\centering
\includegraphics[scale=0.85]{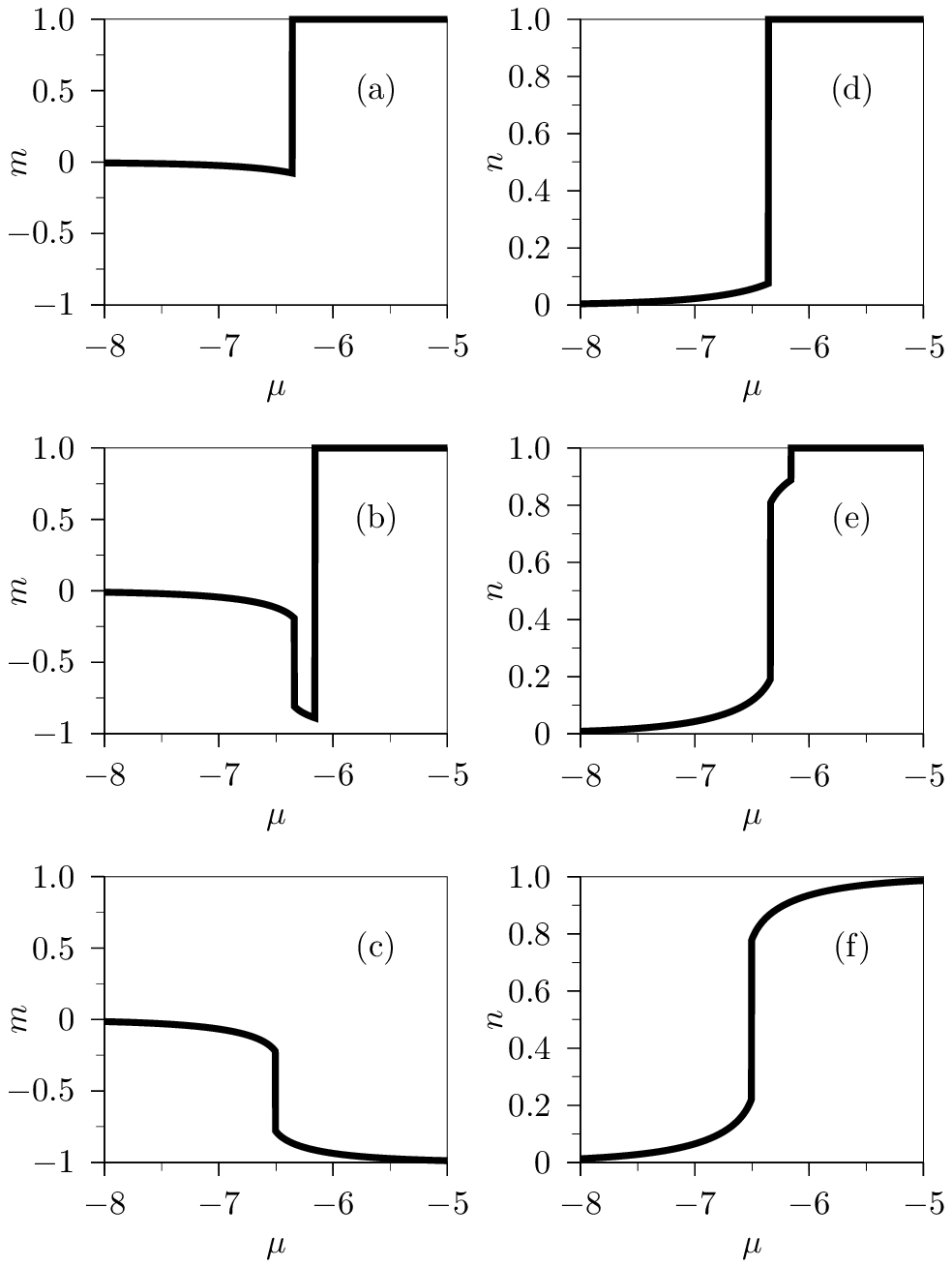}
\caption{(a) Isothermal model transitions with chemical potential $\mu$ are illustrated in Figs (a)-(e),
for temperatures $t$ $0.65, 0.684$ and $0.71$,
in terms of chain order parameter $m$ and density $n$. 
(a) and (d) illustrate the gas-ordered chains transition below the triple point.
(b) and (e) show the double transition, gas-disordered and disordered-ordered
chains, respectively, just above the triple point.  (c) and (f) display data for the
gas-disordered chains transition, further above the triple point. The three cuts
are represented as dashed lines in the phase diagram of Fig. \ref{fig:dia_primeiro}.
}
\label{fig:fig07}
\end{figure}

The different phases and phase transitions present in the phase diagram are
illustrated in Figs. \ref{fig:fig06} and \ref{fig:fig07}. 

In Fig. \ref{fig:fig06},
chain parameter $m$ and lipid density $n$ behaviour with temperature,
at fixed chemical potential,
are shown in different regions of the phase diagram.
The Ord-Dis transition,
with $m$ going from $+1$ to $-1$,
is again shown to be accompanied by a discontinuity in density $n$,
with increasing area per lipid.
However,
the discontinuity in density decreases as chemical potential is raised
(compare Figs. \ref{fig:fig06}(d) and \ref{fig:fig06}(e).
The reentrant behaviour with the sequence of transitions Ord-Gas and Gas-Dis as temperature is raised
is illustrated in Figs. \ref{fig:fig06}(c) and \ref{fig:fig06}(f):
first chain parameter $m$ goes discontinuously from $1$ to $0$,
while density jumps from $1$ to $0$,
and at higher temperature the chain parameter goes from $0$ to $-1$,
while density raises from $0$ to near $80\%$.
The different phase transitions of Fig. \ref{fig:fig06}
are represented as dashed lines in the phase diagram of Fig. \ref{fig:dia_primeiro}.

Chain parameter $m$ and lipid density $n$ behaviour as chemical potential is raised,
at fixed temperature,
is shown in Fig. \ref{fig:fig07},
for different regions of the phase diagram.
The Gas-Ord transition at low temperatures is signaled by a discontinuity in $m$ from $0$ to $1$,
with a density jump from $\approx 0.1$ to $1$ (Figs. \ref{fig:fig07}(a) and \ref{fig:fig07}(d)).
At intermediate temperature,
the Gas-Dis transition is followed by a Dis-Ord transition,
with two discontinuities in density (Figs. \ref{fig:fig07}(b) and \ref{fig:fig07}(d)).
Finally,
at higher temperature,
a Gas-Dis transition is accompanied by a density jump between densities $0.2$ and $0.8$
(Figs. \ref{fig:fig07}(c) and  \ref{fig:fig07}(e) ).
The different phase transitions of Fig. \ref{fig:fig07} are represented as dashed lines in the phase diagram of Fig. \ref{fig:dia_primeiro}.

\begin{figure}
\centering
\includegraphics[scale=0.9]{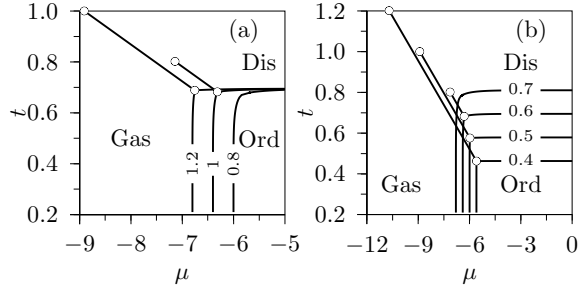}
\caption{Model phase diagrams depend on model parameters. In (a), $\Delta$ is held fixed, while
$K$ is varied (numbers represent values for $K/J$). In (b), $K$ is held fixed, while $\Delta$ 
is varied (numbers represent values for $\Delta/J$). In (a) we see that the gas-order line moves to
higher chemical potentials and the critical point disappears as $K$ is lowered. (b) shows that 
increasing $\Delta$ dislocates both the gas-ordered chains coexistence line as well as the order-disorder 
line, while the critical point disappears.}
\label{fig:fig0teste}
\end{figure}

What is the effect of varying model parameters upon the phase diagram?
Figure \ref{fig:fig0teste}a illustrates the effect of variation of parameter $\Delta$ 
at fixed $K$,
while Fig. \ref{fig:fig0teste}b illustrates the effect of varying $K$ at fixed $\Delta$.
Inspection of role of the interaction parameters in the
expression for energy (Eq. \ref{eq:vh:hamiltoniana}) explain some of the
features displayed by the different phase diagrams.
Parameter $K$ favors site states $\sigma=+1$ and $\sigma=-1$ and thus the
filling of the lattice by lipids, at low temperatures.
Thus, the lipid gas-liquid transition at low temperatures is moved towards lower chemical potential $\mu$,
as $K$ is increased.
On the other hand,
parameter $\Delta$ favors particle state $\sigma=+1$,
and thus stabilizes the the ordered chains liquid state,
moving the low temperature gas-liquid transition to lower chemical potential $\mu$ and
the ordered chain liquid - disordered chain liquid transition to higher temperature $T$.
Dislocation of the coexistence lines might yield the disappearance of the Gas-Dis line,
and therefore of the triple and critical points.
This is the case both for $K=0.8$  ($\Delta=0.6$) in Fig. \ref{fig:fig0teste}a and 
for $\Delta=0.7$ ($K=1$) in Fig. \ref{fig:fig0teste}.

Thus,
the two coexistence lines,
gas liquid and ordered-disordered liquid may either 
merge continuously or meet at a triple point.
But why does the coexistence line 
between the gas and the disordered chain liquid disappear,
as $\Delta$ is increased at fixed $K$ or
as $K$ is increased at fixed $\Delta$?
In fact,
presence of the three phases and of transitions between them 
may be rationalized from analysis of the three limiting models which 
are combined in the Doniach lattice solution we propose: a lattice-gas,
a degenerate lattice-gas and Doniach's model. Analysis of the
phase diagrams of Fig. \ref{fig:fig0teste} in terms of the limiting models is
given in the Appendix.

Finally, we would like to compare Doniach's lattice solution phase diagram with the phase diagram
of the original model, which was given in terms of pressure and temperature.
From thermodynamics, lateral pressure $\Pi$, conjugate to area $A$,
is given by:
\begin{equation}
\Pi = - \frac{1}{A} \Phi(T,A,\mu,H)
\;\text.
\end{equation}
For our model, the thermodynamic grand-potential $\Phi$ is given by Eq. \ref{eq:cm_f}.

\begin{figure}
\includegraphics[scale=0.9]{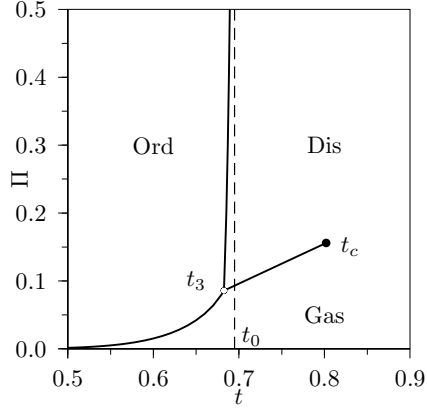}
\caption{Lateral pressure {\it vs} versus temperature. 
Dashed line indicates the high density limit 
 of the order-disorder transition, temperature $t_o$.}
\label{fig:tp}
\end{figure}

Figure  \ref{fig:tp} displays the lateral pressure $\Pi$ versus temperature $t$ phase diagram,
for the same model parameters as in Fig. \ref{fig:dia_primeiro}.
The gas phase is present at low pressure and higher temperatures,
as for usual fluids.
At higher pressures,
the two fluid phases are separated by a coexistence line,
with the ordered chain liquid (Ord) at lower temperatures,
and the disordered chains liquid (Dis) at higher temperatures.
The coexistence pressure $\Pi^*$ between the two liquid phases
rises steeply as temperature is increased.

In the case of the  model by Doniach, 
pressure is a linear function of temperature
at the order-disorder coexistence line,
which ends at a critical point.
In that case,
a unique order parameter is present,
with $m(a)$ (see Eq. \ref{eq:a(m)})
and the chain order-disorder transition accompanies the
pseudo -density transition. In the  model we propose,
the chain order-disorder transition is associated to a true density transition.

At low chemical potential and pressure, chain disordering is accompanied
by a density gap, which goes exponentially to zero at higher potentials.
The linear dependence between coexistence temperature and 'pressure' disappears.

\section{Physical interpretation: monolayers {\it vs} bilayers}

In order to interpret our findings in terms of the two systems of interest, monolayers
and bilayers, we must analyse the differences between inter-particle interactions in 
the two systems, as well as the physical boundary conditions involved. 

The anisotropic organization of the phospholipid molecules in layers 
is a consequence of the fact that those are amphiphilic molecules, with a hydrophylic polar headgroup
which mixes with water and a hydrophobic hydrocarbonic tail which would phase separate were it
not attached to the polar headgroup. This nematic like structure is common to bilayers and
monolayers. However, there are specific aspects of the {\em interactions} between lipids and 
water which make them different physical systems. While headgroups are in contact with water
in both systems, hydrocarbonic tails turn to the air subphase and have no contact with water, 
independently of the distance between lipids,
 in the case of monolayers, whereas for bilayers hydrophobic chains turn to the hydrophobic bilayer
core, but water penetration increases as lipid molecules go apart. 

Our model system is a plane system in two dimensions.
Monolayers are truly two-dimensional,
while bilayers may be seen as two weakly interacting spherical monolayers.
Monolayers reside in the interface between water and air,
while bilayers are in bulk water,
albeit each leaflet of the bilayer could be taken as located on a water-hydrocarbon interface.
Monolayers may be manipulated both through direct compression,
as well as through heating,
implying a line of  disordering transitions in the pressure-temperature plane. 
Bilayer behaviour is probed through temperature variations only,
and the disordering transition occurs at a single temperature.

In the following subsections we analyse the differences pointed and out above and the relation to our model.

\subsection{Monolayers}

A monolayer is constituted by lipid particles,
whose chains suffer Van der Waals attraction.
Lipid headgroups rest on the water surface,
and lipid chains do not get in contact with water molecules,
which allows us to  take null lipid-water interaction parameters,
$\epsilon_\text{lip,w} = \epsilon_\text{ow}=\epsilon_\text{dw}=0$. 
On the other hand,
water-water ``bonds'' are surface ``bonds'',
from now on labelled as $\epsilon_\text{ww}^\text{surf}$.
When lipid molecules become dispersed,
water molecules attract strongly between themselves yielding large surface 
tension.
At very low temperature and very low pressure, 
one expects the model system to go into a 'gas' lipid phase,
since water-water interactions are dominant
over lipid-lipid interactions.

\begin{figure}
\includegraphics[scale=0.9]{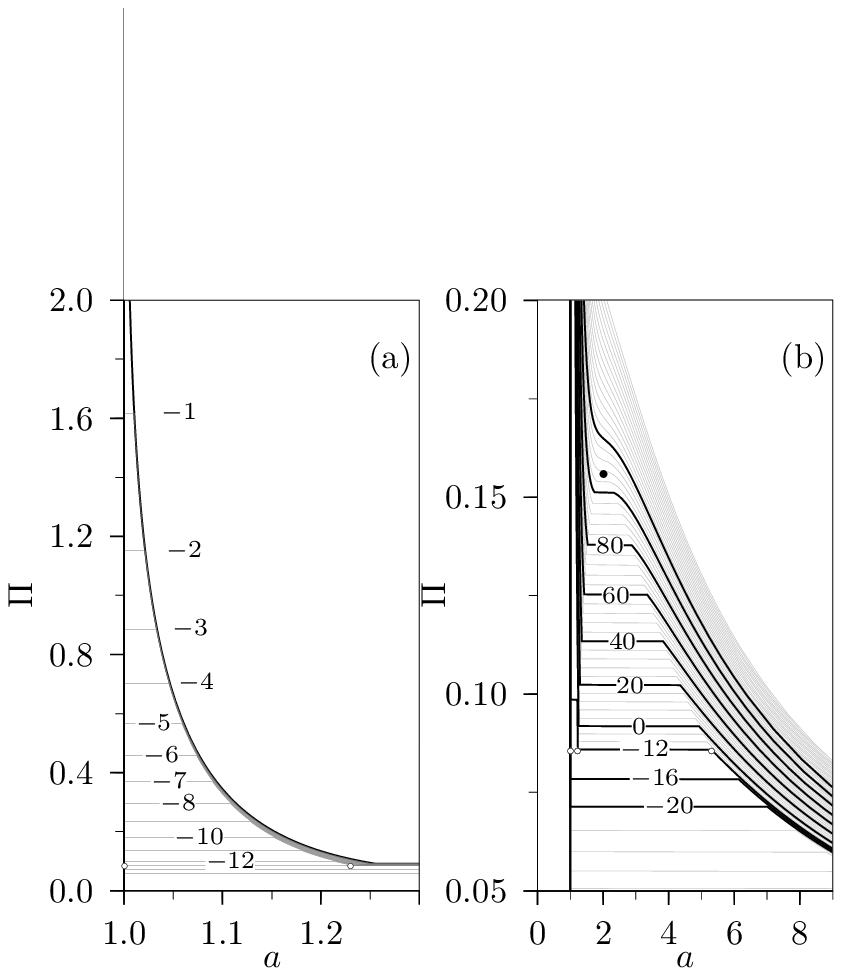}
\caption{Pressure {\it vs} lipid area isotherms.
 $\Delta=0.6$, $K=1$ and $\Omega=1000$.
Figures indicate temperatures referred to the high density limit of the order-disorder transition,
$t_0$ (see Fig 9) as $(t-t_0)*1000$. Isotherms present two plateaus
corresponding to the order-disorder and the gas-liquid transitions, for
$t_3<t<t_0$.
Isotherms span a large range of areas, and are thus represented
separately for the two transitions. (a) On the left, pressures are higher and area variation
remains within 30\% upon the order-disorder transition. The order-disorder transition
is present only below $t_0$. (b) On the right, 
pressures are an order of magnitude lower, and area may vary by a factor of 10, 
along the gas-liquid disordered liquid above $t_3$ ($t_0 - 12/1000$)
and gas-ordered liquid below $t_3$. 
}
\label{fig:isotermas}
\end{figure}

Much of the experimental investigation of monolayers 
is given in terms of Langmuir pressure-area isotherms,
which present two coexistence plateaus,
one between the gas and the expanded liquid, at lower pressure,
and the second one between the expanded and the condensed liquid phases
\cite{mohwald1995,kaganer1999}, for which the discontinuity in area per molecule
is an order of magnitude lower.
If compression is further increased, collapse of the monolayer
comes about \cite{lee2008}.
Our model isotherms displayed in Fig. \ref{fig:isotermas} compare well qualitatively
to experimental plots \cite{marcelja1979}.
Both transitions are present,
at different orders of magnitude both for pressure and area per molecule.
The lattice, of course,
limits the minimum area, so that isotherms increase steeply at the lower limit,
differently from the experimental system,
which,
besides,
may be allowed to expand into a 3rd dimension.  (see Figs. \ref{fig:fig06} an \ref{fig:fig07}d-e)

A critical point for the ord-dis transition is absent for the parameters explored
in this study.
However,
it may be present for a different set of parameters,
as explained in the appendix. 

%

\subsection {Bilayers }

In the case of a bilayer,
the gas-liquid transition would correspond to 
membrane disaggregation and some critical micellar parameter \cite{shida1998},
which is not of interest in the study of biomembranes.
As for the integral vesicle thermal phases,
differently from the monolayer case,
water-lipid interactions are essential: 
they are the source of the ``hydrophobic'' interaction,
$\epsilon_\text{hydroph} \equiv \epsilon_\text{ww}^\text{bulk}-\epsilon_\text{lw}$,
with
$\epsilon_\text{ow} \ne 0$ and $ \epsilon_\text{dw} \ne 0$.
Also,
water-water ``bonds'' in the bulk are ``looser'' than at the surface,
and we thus label them as $\epsilon_\text{ww}^\text{bulk}$.

In relation to pressure effect,
bilayers may be considered to be in a tension-free state,
which corresponds \cite{marsh1996, nagle2000, gruen1982} to a situation of zero 
lateral pressure $\Pi$.  Different authors propose an equivalence at some specific
lateral pressures, but lately this equivalence has been questioned.

\subsection {Bilayers {\it vs} monolayers}

How then are we to associate the model phase diagrams,
Fig. \ref{fig:dia_primeiro} and Fig. \ref{fig:tp}, 
to physical monolayers and bilayers?

A reasonable simplification is to take lipid-water interactions independent of lipid state,
with
$\epsilon_\text{ow}=\epsilon_\text{dw}=\epsilon_\text{lw}$.
This assumption yields the following relations between monolayer and bilayer parameters 
(see Eqns. \ref{eq:parameters:J} - \ref{eq:parameters:I}):
\begin{equation}
J^\text{M}=J^\text{B}
\;\text,
\end{equation}
\begin{equation}
\Delta^\text{M}=\Delta^\text{B}
\;\text,
\end{equation}
\begin{equation}
K^\text{M} =
K^\text{B}
+ (\epsilon_\text{ww}^\text{surf} - \epsilon_\text{ww}^\text{bulk}) 
+2 \epsilon_\text{lw}
\;\text,
\label{eq:KmKb}
\end{equation}
and
\begin{equation}
I^\text{M} =
I^\text{B}
+ 4(\epsilon_\text{ww}^\text{surf} - \epsilon_\text{ww}^\text{bulk}) 
+ 4 \epsilon_\text{lw}
\;\text.
\end{equation}

What are the implications of the difference in energy parameters for the two systems?

Let us first analyse the differences between parameters $K^\text{M}$ and $K^\text{B}$. 
Interactions between water molecules on the surface are more stable than
between molecules in the bulk, which implies
$\epsilon_\text{ww}^\text{surf} > \epsilon_\text{ww}^\text{bulk}$.
Thus we have $K^\text{M} > K^\text{B}$.
This is an important result, since it implies that data for the
two systems cannot be mapped onto the same phase diagram. 
We inspect Fig. \ref{fig:fig0teste}a taking into account this point.
Variation of $K$ has two simultaneous effects.
As it increases, (i) it displaces the gas-liquid line to lower chemical potential,
as should be expected, since it favours lipid-lipid interactions; (ii) it turns the 
order-disorder coexistence line near the triple point less dependent on temperature.

The last effect has implications on the area discontinuity upon the order-disorder 
transition. If one focuses on this transition at some specific temperature, the discontinuity in
area is smaller for larger $K$. This means that, if effective chemical potential $\mu_\text{eff}$ is
the same for both systems, adopting the same transition temperature implies obtaining
different areas for the two systems at coexistence. Interestingly, this is what happens
with the experimental systems: if one looks for the equivalence at the same transition 
temperature, areas are different. If, on the other hand, equivalence is sought for 
from the same area gap, the two transitions are found at different temperatures.

A further difference between the two systems arises if we try inspection of the 
pressure-temperature phase diagram (Fig. \ref{fig:tp}).
Different suggestions can be found in the literature for the lateral pressure on monolayers which would render
them equivalent to bilayers \cite{marsh1996, nagle2000}.
If,
for the sake of further analysing the pursuit of equivalence between the two systems,
we ignore the differences in $K$,
it is possible to establish a relation between the lateral pressures of the two systems.
We consider a particular thermodynamic state for the model system,
which corresponds to a point in the $\mu$ {\it vs} $t$ phase diagram,
and to the same order parameter values for $m$ and $n$ (Eqns. \ref{eq:cm_mn}).
Note that if the two systems,
monolayer and bilayer are considered to be in the same thermodynamic system, 
$m^\text{B}=m^\text{M}$ and $n^\text{B}=n^\text{M}$,
besides having equal $K$ parameters,
both systems must have equal effective chemical potentials  $\mu$,
{\it i.e.} $\mu^\text{M}=\mu^\text{B}$.
What are the implications on lateral pressure?


Lateral pressure is related to the grand-potential through Eq. \ref{eq:cm_mn}. 
From the definition of the grand potential \cite{callen}, 

\begin{eqnarray}
\Phi & = & \left< E \right> - TS - \mu_\text{l} N_\text{l} - \mu_\text{w} N_\text{w} ) \nonumber \\
& = & \left< E_\text{int} - TS - \mu N_\text{l} - (\mu_\text{w}+2\epsilon_\text{ww}) A \right>
\end{eqnarray}

For the same thermodynamic 
state, for $K^M=K^B$,
the grand-potential for the two systems will differ only through the  constant terms,
since density of lipids and entropy must be the same.
Thus
\begin{equation}
\Phi^\text{B}-\Phi^\text{M} = 
- \left( K^\text{B} - K^\text{M} \right) N_\text{ll} 
-2 \left( \epsilon_\text{ww}^\text{bulk} - \epsilon_\text{ww}^\text{surf} \right) A
\;\text,
\end{equation}

which yields the following simple relation for
the lateral pressures of the two systems:
\begin{equation}
\Pi^\text{M} -\Pi^\text{B} =
2 \left( 
\epsilon_\text{ww}^\text{surf}-\epsilon_\text{ww}^\text{bulk}
\right)
\;\text,
\end{equation}
since  $\Phi=-\Pi A$. 

For the null pressure of the bilayer,
corresponds a positive lateral pressure on the monolayer,
\begin{equation}
\Pi^\text{M} =
2( \epsilon_\text{ww}^\text{surf} - \epsilon_\text{ww}^\text{bulk})
\;\text,
\end{equation}
since $\epsilon_\text{ww}^\text{surf} > \epsilon_\text{ww}^\text{bulk}$.

This result is qualitatively in agreement with several proposals of the literature based on
experimental measurements \cite{nagle1976,gruen1982, marsh1996} and gives it
an interpretation in terms of the statistical model.
In particular,
it is also in line with a phenomenological analysis proposed by Marsh \cite{marsh1996},
in which the monolayer pressure corresponding to the membrane thermodynamic state
at the main transition would be numerically equal to the
hydrophobic free energy density. The origin of the difference in pressures would be the water surface tension, 
a consequence of ``stronger'' hydrogen bonds on the surface,
as compared to bulk water.

\begin{figure}
\includegraphics[scale=0.9]{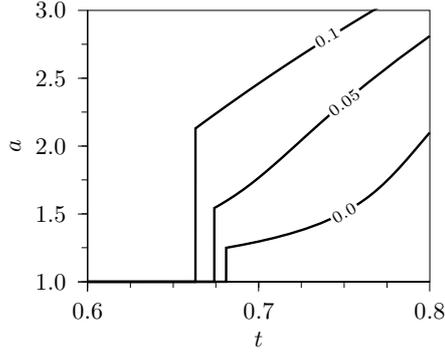}
\caption{Area per lipid as function of temperature at fixed
pressure for the order-disorder transition. $K/J=1$
and $\Delta/J=0.6$.
$\Pi^\text{M} = 0.1$ or $\Pi^\text{B}=0$ (see text) for $K^\text{M}=K^\text{B}$
and values of $\epsilon_\text{lw}/J$} are as indicated.
Density gap decreases as mixing with water is favoured.
\label{fig:area}
\end{figure}

Figure \ref{fig:area} illustrates the behaviour of the density gap for different
ratios of the lipid-water interaction to water surface tension, at fixed pressure,
under the artificial condition of equivalence ($K^\text{M}=K^\text{B}$).
Increasing lipid-water interaction,
with respect to water surface tension,
transition temperature is decreased, 
while the area discontinuity increases.

However, the result for the equivalence pressure just presented, as stated before, 
is based on the artificial assumption of equal interaction
coefficient $K$ for the two systems. But Eq. \ref{eq:KmKb} shows that this assumption is inconsistent
with the presence of either water-water or water-lipid interactions. Therefore, the difference
in the interaction strength of water molecules on the surface and in bulk, as well as the 
presence of water-lipid interactions only for the bulk lipid layer make inconsistent
the hypothesis of equivalence between monolayers and bilayers, and explains the
difficulty in aligning simultaneously both the transition temperature and the density gap
\cite{nagle2000}.

\section{Final comments}

We have proposed a generalization of the two-state model for lipid layers,
which allows an exact description of local density. 
This is essential for the investigation
of the effect of charges, in the case of dissociating lipids.

Inspection of model properties with respect to the relation
between density and chain order led to some additional
conclusions: 

(i) for monolayers, the model describes both the
liquid transitions (order-disorder or condensed liquid-expanded liquid),
in good qualitative agreement with experimental studies; 

(ii) analysis in terms of the different model interactions between lipids and water for
monolayers and bilayers yields an explanation for
the difficulty in establishing the equivalence between the two
experimental systems.

Further investigation of the model system in the presence of charges, both for
dissociating headgroups as well as for dipolar headgroups, is underway.

\section{Acknowledgments}

We thank Eduardo Henriques at UFPEL for pointing out the possibility
of adapting our model to the study
of monolayers and to Mario Tamashiro at UNICAMP for many conversations
on the theme of our work.

\appendix
\section*{APPENDIX: Limiting models}

Our model may be thought as a composition of three models:
(i) the order-disorder Doniach model, of density {\it one},
(ii) a simple lattice gas,
and (iii) a degenerate lattice gas. 
The three limiting models are 
obtained if one of the three values for
site variables $\sigma$ is discarded.
Model (i) results from making site variables $\sigma$ equal to $1$ and $-1$.
Model (ii) results from restricting site variables $\sigma$ to $0$ and $+1$. 
Model (iii) is obtained if  site variables $\sigma$ are taken as $0$ and $-1$.
Under such restrictions,
each one of the three limiting models may be mapped on the two-state Ising model,
given by
\begin{equation}
E_\text{Ising}=
-\tilde J \sum_{(ij)} s_i s_j
-\tilde H \sum_{i} s_i
\;\text,
\end{equation}
with particle states $s$ assuming two possible values,
$s=+1$  or $s=-1$. 
Model parameters $\tilde J$ and $\tilde H$ are different for each model,
and if given in terms of the Doniach lattice gas parameters (Eqs. 
\ref{eq:parameters:J} 
\ref{eq:parameters:Delta} 
\ref{eq:parameters:K} 
),
are as follows:
For model (i),
\begin{equation}
\tilde J_\text{i} = J
\;\text;
\end{equation}
\begin{equation}
\tilde H_\text{i}(T) = 4 \Delta 
- \frac{1}{2 \beta} \ln \Omega
\;\text.
\label{eq:analitico:h1}
\end{equation}
For model (ii)
\begin{equation}
\tilde J_\text{ii} = \frac14 (J + 2\Delta + K)
\;\text;
\end{equation}
\begin{equation}
\tilde H_\text{ii} (\mu)= J + 2 \Delta + K + \frac12 \mu
\;\text.
\end{equation}
Finally, for model (iii), we have
\begin{equation}
\tilde J_\text{iii} = \frac14 ( J - 2 \Delta + K)
\;\text;
\end{equation}
and
\begin{equation}
\tilde H_\text{iii}(T,\mu) = J + K - 2 \Delta + \frac12 \mu
+ \frac{1}{2 \beta} \ln \Omega
\;\text.
\label{eq:analitico:h3}
\end{equation}

The phase behaviour of each of the three models may be obtained by
adapting well-known results for the Ising model for ferromagnetism.
The Ising model presents a coexistence line at $\tilde H=0$ which ends at a critical
temperature $T_C$, given by 
\begin{equation}
\frac{k_B T_C}{\tilde J}=4
\;\text.
\end{equation}

As a result,
a coexistence line and a critical point exists for each of the three limiting models.
Model (i) displays a coexistence line at fixed temperature
\begin{equation}
t_0=\frac{8 \Delta}{J \ln \Omega}
\label{eq:tzero}
\end{equation}
between the disordered ($m=1$) and ordered chain ($m=-1$) liquids.  
Model (ii) presents coexistence
at fixed chemical potential 
\begin{equation}
\mu_0= - 2(J+2 \Delta + K)
\;\text,
\end{equation} 
between a gas ($n=0$) and a simple liquid ($n>0$), 
with a critical point at $t_{C, (\text{ii})} = 1+2 \Delta/J + K/J $. 
Finally,
in model (iii) a gas ($n=0$) and a degenerate liquid ($n>0$) coexist at
$\mu(t)=  - 2(J - 2 \Delta + K) - k_B T \ln \Omega$,
with a critical point at 
\begin{equation}
t_{C, (\text{ii})} = 1 - 2 \frac{\Delta}{J}+ \frac{K}{J}
\;\text.
\end{equation}

\begin{figure}
\includegraphics[scale=0.9]{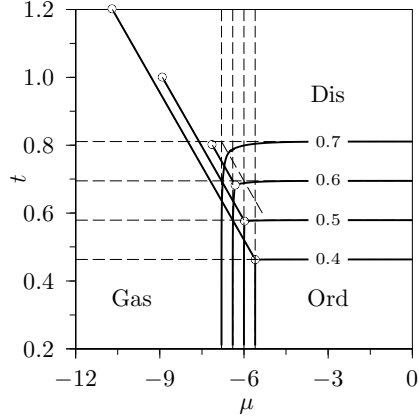}
\caption{Coexistence lines for DLG (symbols) and coexistence lines for
the limiting models: Doniach's model (horizontal dashed lines), lattice gas (vertical 
dashed lines) and degenerate lattice gas (sloped dashed lines). Critical points
for the limiting models are seen only for the case of the degenerate lattice gas.
For the other two limiting systems, critical points are outside frame. For
$\Delta=0.7$, the DLG presents no Gas-Dis line. Also, the critical temperature  
Gas-Liquid line  of the limiting degenerate
lattice gas model is below the temperature of the Ord-Dis of the limiting
Doniach model. }
\label{fig:dia_analitico}
\end{figure}

In Fig. \ref{fig:dia_analitico} we compare phase coexistence lines 
of previous Fig. \ref{fig:fig0teste}b
with coexistence lines for the limiting models (i)-(iii), at different
values of $\Delta$.
As can be seen,
for the case in which the critical temperature of the limiting model 
(ii) is lower than the temperature for the chain order-disorder transition the line G-Dis disappears,
as expected.
Similar analysis explain previous Fig. \ref{fig:fig0teste} a.
The analysis of the limiting models also allows
us to expect a critical point at the end of the Ord-Dis coexistence line 
if $t_0>4$.

\bibliography{mybib}

\begin{thebibliography}{28}%
\makeatletter
\providecommand \@ifxundefined [1]{%
 \@ifx{#1\undefined}
}%
\providecommand \@ifnum [1]{%
 \ifnum #1\expandafter \@firstoftwo
 \else \expandafter \@secondoftwo
 \fi
}%
\providecommand \@ifx [1]{%
 \ifx #1\expandafter \@firstoftwo
 \else \expandafter \@secondoftwo
 \fi
}%
\providecommand \natexlab [1]{#1}%
\providecommand \enquote  [1]{``#1''}%
\providecommand \bibnamefont  [1]{#1}%
\providecommand \bibfnamefont [1]{#1}%
\providecommand \citenamefont [1]{#1}%
\providecommand \href@noop [0]{\@secondoftwo}%
\providecommand \href [0]{\begingroup \@sanitize@url \@href}%
\providecommand \@href[1]{\@@startlink{#1}\@@href}%
\providecommand \@@href[1]{\endgroup#1\@@endlink}%
\providecommand \@sanitize@url [0]{\catcode `\\12\catcode `\$12\catcode
  `\&12\catcode `\#12\catcode `\^12\catcode `\_12\catcode `\%12\relax}%
\providecommand \@@startlink[1]{}%
\providecommand \@@endlink[0]{}%
\providecommand \url  [0]{\begingroup\@sanitize@url \@url }%
\providecommand \@url [1]{\endgroup\@href {#1}{\urlprefix }}%
\providecommand \urlprefix  [0]{URL }%
\providecommand \Eprint [0]{\href }%
\providecommand \doibase [0]{http://dx.doi.org/}%
\providecommand \selectlanguage [0]{\@gobble}%
\providecommand \bibinfo  [0]{\@secondoftwo}%
\providecommand \bibfield  [0]{\@secondoftwo}%
\providecommand \translation [1]{[#1]}%
\providecommand \BibitemOpen [0]{}%
\providecommand \bibitemStop [0]{}%
\providecommand \bibitemNoStop [0]{.\EOS\space}%
\providecommand \EOS [0]{\spacefactor3000\relax}%
\providecommand \BibitemShut  [1]{\csname bibitem#1\endcsname}%
\let\auto@bib@innerbib\@empty
\bibitem [{\citenamefont {Kaganer}\ \emph {et~al.}(1999)\citenamefont
  {Kaganer}, \citenamefont {M{\"{o}}hwald},\ and\ \citenamefont
  {Dutta}}]{kaganer1999}%
  \BibitemOpen
  \bibfield  {author} {\bibinfo {author} {\bibfnamefont {V.~M.}\ \bibnamefont
  {Kaganer}}, \bibinfo {author} {\bibfnamefont {H.}~\bibnamefont
  {M{\"{o}}hwald}}, \ and\ \bibinfo {author} {\bibfnamefont {P.}~\bibnamefont
  {Dutta}},\ }\href@noop {} {\bibfield  {journal} {\bibinfo  {journal} {Rev.
  Mod. Phys.}\ }\textbf {\bibinfo {volume} {71}},\ \bibinfo {pages} {779}
  (\bibinfo {year} {1999})}\BibitemShut {NoStop}%
\bibitem [{\citenamefont {M{\"{o}}hwald}(1995)}]{mohwald1995}%
  \BibitemOpen
  \bibfield  {author} {\bibinfo {author} {\bibfnamefont {H.}~\bibnamefont
  {M{\"{o}}hwald}},\ }in\ \href@noop {} {\emph {\bibinfo {booktitle} {Structure
  and Dynamics of Membranes}}},\ Vol.~\bibinfo {volume} {1},\ \bibinfo {editor}
  {edited by\ \bibinfo {editor} {\bibfnamefont {R.}~\bibnamefont {Lipowsky}}\
  and\ \bibinfo {editor} {\bibfnamefont {E.}~\bibnamefont {Sackmann}}}\
  (\bibinfo  {publisher} {North-Holland},\ \bibinfo {year} {1995})\ Chap.\
  \bibinfo {chapter} {Handbook of Biological Physics}, pp.\ \bibinfo {pages}
  {161--211}\BibitemShut {NoStop}%
\bibitem [{\citenamefont {Bloom}\ \emph {et~al.}(1991)\citenamefont {Bloom},
  \citenamefont {Evans},\ and\ \citenamefont {Mouritsen}}]{bloom1991}%
  \BibitemOpen
  \bibfield  {author} {\bibinfo {author} {\bibfnamefont {M.}~\bibnamefont
  {Bloom}}, \bibinfo {author} {\bibfnamefont {E.}~\bibnamefont {Evans}}, \ and\
  \bibinfo {author} {\bibfnamefont {O.~G.}\ \bibnamefont {Mouritsen}},\
  }\href@noop {} {\bibfield  {journal} {\bibinfo  {journal} {Quarterly reviews
  of biophysics}\ }\textbf {\bibinfo {volume} {24}},\ \bibinfo {pages} {293}
  (\bibinfo {year} {1991})}\BibitemShut {NoStop}%
\bibitem [{\citenamefont {Tristram-Nagle}\ and\ \citenamefont
  {Nagle}(2004)}]{nagle2004}%
  \BibitemOpen
  \bibfield  {author} {\bibinfo {author} {\bibfnamefont {S.}~\bibnamefont
  {Tristram-Nagle}}\ and\ \bibinfo {author} {\bibfnamefont {J.~F.}\
  \bibnamefont {Nagle}},\ }\href@noop {} {\bibfield  {journal} {\bibinfo
  {journal} {Chemistry and Physics of Lipids}\ }\textbf {\bibinfo {volume}
  {127}},\ \bibinfo {pages} {3} (\bibinfo {year} {2004})}\BibitemShut {NoStop}%
\bibitem [{\citenamefont {Nagle}(1973)}]{nagle1973}%
  \BibitemOpen
  \bibfield  {author} {\bibinfo {author} {\bibfnamefont {J.~F.}\ \bibnamefont
  {Nagle}},\ }\href@noop {} {\bibfield  {journal} {\bibinfo  {journal} {The
  Journal of Chemical Physics}\ }\textbf {\bibinfo {volume} {58}},\ \bibinfo
  {pages} {252} (\bibinfo {year} {1973})}\BibitemShut {NoStop}%
\bibitem [{\citenamefont {Nagle}(1975)}]{nagle1975}%
  \BibitemOpen
  \bibfield  {author} {\bibinfo {author} {\bibfnamefont {J.~F.}\ \bibnamefont
  {Nagle}},\ }\href@noop {} {\bibfield  {journal} {\bibinfo  {journal} {The
  Journal of Chemical Physics}\ }\textbf {\bibinfo {volume} {63}},\ \bibinfo
  {pages} {1255} (\bibinfo {year} {1975})}\BibitemShut {NoStop}%
\bibitem [{\citenamefont {Marcelja}(1974)}]{marcelja1974}%
  \BibitemOpen
  \bibfield  {author} {\bibinfo {author} {\bibfnamefont {S.}~\bibnamefont
  {Marcelja}},\ }\href@noop {} {\bibfield  {journal} {\bibinfo  {journal}
  {Biochimica et Biophysica Acta (BBA) - Biomembranes}\ }\textbf {\bibinfo
  {volume} {367}},\ \bibinfo {pages} {165} (\bibinfo {year}
  {1974})}\BibitemShut {NoStop}%
\bibitem [{\citenamefont {Nagle}(1976)}]{nagle1976}%
  \BibitemOpen
  \bibfield  {author} {\bibinfo {author} {\bibfnamefont {J.~F.}\ \bibnamefont
  {Nagle}},\ }\href@noop {} {\bibfield  {journal} {\bibinfo  {journal} {The
  Journal of Membrane Biology}\ }\textbf {\bibinfo {volume} {27}},\ \bibinfo
  {pages} {233} (\bibinfo {year} {1976})}\BibitemShut {NoStop}%
\bibitem [{\citenamefont {Gruen}\ and\ \citenamefont
  {Wolfe}(1982)}]{gruen1982}%
  \BibitemOpen
  \bibfield  {author} {\bibinfo {author} {\bibfnamefont {D.~W.~R.}\
  \bibnamefont {Gruen}}\ and\ \bibinfo {author} {\bibfnamefont
  {J.}~\bibnamefont {Wolfe}},\ }\href@noop {} {\bibfield  {journal} {\bibinfo
  {journal} {Biochimica et Biophysica Acta (BBA) - Biomembranes}\ }\textbf
  {\bibinfo {volume} {688}},\ \bibinfo {pages} {572} (\bibinfo {year}
  {1982})}\BibitemShut {NoStop}%
\bibitem [{\citenamefont {Marsh}(1996)}]{marsh1996}%
  \BibitemOpen
  \bibfield  {author} {\bibinfo {author} {\bibfnamefont {D.}~\bibnamefont
  {Marsh}},\ }\href@noop {} {\bibfield  {journal} {\bibinfo  {journal}
  {Biochimica et Biophysica Acta (BBA) - Reviews on Biomembranes}\ }\textbf
  {\bibinfo {volume} {1286}},\ \bibinfo {pages} {183} (\bibinfo {year}
  {1996})}\BibitemShut {NoStop}%
\bibitem [{\citenamefont {Ni}\ \emph {et~al.}(2013)\citenamefont {Ni},
  \citenamefont {Gruenbaum},\ and\ \citenamefont {Skinner}}]{ni2013}%
  \BibitemOpen
  \bibfield  {author} {\bibinfo {author} {\bibfnamefont {Y.}~\bibnamefont
  {Ni}}, \bibinfo {author} {\bibfnamefont {S.~M.}\ \bibnamefont {Gruenbaum}}, \
  and\ \bibinfo {author} {\bibfnamefont {J.~L.}\ \bibnamefont {Skinner}},\
  }\href@noop {} {\bibfield  {journal} {\bibinfo  {journal} {Proceedings of the
  National Academy of Sciences}\ }\textbf {\bibinfo {volume} {110}},\ \bibinfo
  {pages} {1992} (\bibinfo {year} {2013})}\BibitemShut {NoStop}%
\bibitem [{\citenamefont {Caille}\ \emph {et~al.}(1978)\citenamefont {Caille},
  \citenamefont {Rapini}, \citenamefont {Zuckermann}, \citenamefont {Cros},\
  and\ \citenamefont {Doniach}}]{caille1978}%
  \BibitemOpen
  \bibfield  {author} {\bibinfo {author} {\bibfnamefont {A.}~\bibnamefont
  {Caille}}, \bibinfo {author} {\bibfnamefont {A.}~\bibnamefont {Rapini}},
  \bibinfo {author} {\bibfnamefont {M.~J.}\ \bibnamefont {Zuckermann}},
  \bibinfo {author} {\bibfnamefont {A.}~\bibnamefont {Cros}}, \ and\ \bibinfo
  {author} {\bibfnamefont {S.}~\bibnamefont {Doniach}},\ }\href@noop {}
  {\bibfield  {journal} {\bibinfo  {journal} {Canadian Journal of Physics}\
  }\textbf {\bibinfo {volume} {56}},\ \bibinfo {pages} {348} (\bibinfo {year}
  {1978})}\BibitemShut {NoStop}%
\bibitem [{\citenamefont {Doniach}(1978)}]{doniach1978}%
  \BibitemOpen
  \bibfield  {author} {\bibinfo {author} {\bibfnamefont {S.}~\bibnamefont
  {Doniach}},\ }\href@noop {} {\bibfield  {journal} {\bibinfo  {journal} {The
  Journal of Chemical Physics}\ }\textbf {\bibinfo {volume} {68}},\ \bibinfo
  {pages} {4912} (\bibinfo {year} {1978})}\BibitemShut {NoStop}%
\bibitem [{\citenamefont {Mouritsen}\ \emph {et~al.}(1983)\citenamefont
  {Mouritsen}, \citenamefont {Boothroyd}, \citenamefont {Harris}, \citenamefont
  {Jan}, \citenamefont {Lookman}, \citenamefont {MacDonald}, \citenamefont
  {Pink},\ and\ \citenamefont {Zuckermann}}]{mouritsen1983}%
  \BibitemOpen
  \bibfield  {author} {\bibinfo {author} {\bibfnamefont {O.~G.}\ \bibnamefont
  {Mouritsen}}, \bibinfo {author} {\bibfnamefont {A.}~\bibnamefont
  {Boothroyd}}, \bibinfo {author} {\bibfnamefont {R.}~\bibnamefont {Harris}},
  \bibinfo {author} {\bibfnamefont {N.}~\bibnamefont {Jan}}, \bibinfo {author}
  {\bibfnamefont {T.}~\bibnamefont {Lookman}}, \bibinfo {author} {\bibfnamefont
  {L.}~\bibnamefont {MacDonald}}, \bibinfo {author} {\bibfnamefont {D.~A.}\
  \bibnamefont {Pink}}, \ and\ \bibinfo {author} {\bibfnamefont {M.~J.}\
  \bibnamefont {Zuckermann}},\ }\href@noop {} {\bibfield  {journal} {\bibinfo
  {journal} {The Journal of Chemical Physics}\ }\textbf {\bibinfo {volume}
  {79}},\ \bibinfo {pages} {2027} (\bibinfo {year} {1983})}\BibitemShut
  {NoStop}%
\bibitem [{\citenamefont {Baret}\ and\ \citenamefont
  {Firpo}(1983)}]{baret1983}%
  \BibitemOpen
  \bibfield  {author} {\bibinfo {author} {\bibfnamefont {J.-F.}\ \bibnamefont
  {Baret}}\ and\ \bibinfo {author} {\bibfnamefont {J.-L.}\ \bibnamefont
  {Firpo}},\ }\href@noop {} {\bibfield  {journal} {\bibinfo  {journal} {Journal
  of Colloid and Interface Science}\ }\textbf {\bibinfo {volume} {94}},\
  \bibinfo {pages} {487} (\bibinfo {year} {1983})}\BibitemShut {NoStop}%
\bibitem [{\citenamefont {Pink}\ and\ \citenamefont
  {Chapman}(1979)}]{pink1979}%
  \BibitemOpen
  \bibfield  {author} {\bibinfo {author} {\bibfnamefont {D.~A.}\ \bibnamefont
  {Pink}}\ and\ \bibinfo {author} {\bibfnamefont {D.}~\bibnamefont {Chapman}},\
  }\href@noop {} {\bibfield  {journal} {\bibinfo  {journal} {Proceedings of the
  National Academy of Sciences}\ }\textbf {\bibinfo {volume} {76}},\ \bibinfo
  {pages} {1542} (\bibinfo {year} {1979})}\BibitemShut {NoStop}%
\bibitem [{\citenamefont {Heimburg}(2007)}]{heimburg}%
  \BibitemOpen
  \bibfield  {author} {\bibinfo {author} {\bibfnamefont {T.}~\bibnamefont
  {Heimburg}},\ }\href@noop {} {\emph {\bibinfo {title} {Thermal Biophysics of
  Membranes (Tutorials in Biophysics)}}},\ \bibinfo {edition} {1st}\ ed.\
  (\bibinfo  {publisher} {Wiley-VCH},\ \bibinfo {year} {2007})\BibitemShut
  {NoStop}%
\bibitem [{\citenamefont {Almeida}(2011)}]{almeida2011}%
  \BibitemOpen
  \bibfield  {author} {\bibinfo {author} {\bibfnamefont {P.~F.}\ \bibnamefont
  {Almeida}},\ }\href@noop {} {\bibfield  {journal} {\bibinfo  {journal}
  {Biophysical journal}\ }\textbf {\bibinfo {volume} {100}},\ \bibinfo {pages}
  {420} (\bibinfo {year} {2011})}\BibitemShut {NoStop}%
\bibitem [{\citenamefont {Tamashiro}\ \emph {et~al.}(2011)\citenamefont
  {Tamashiro}, \citenamefont {Barbetta}, \citenamefont {Germano},\ and\
  \citenamefont {Henriques}}]{tamashiro2011}%
  \BibitemOpen
  \bibfield  {author} {\bibinfo {author} {\bibfnamefont {M.~N.}\ \bibnamefont
  {Tamashiro}}, \bibinfo {author} {\bibfnamefont {C.}~\bibnamefont {Barbetta}},
  \bibinfo {author} {\bibfnamefont {R.}~\bibnamefont {Germano}}, \ and\
  \bibinfo {author} {\bibfnamefont {V.~B.}\ \bibnamefont {Henriques}},\
  }\href@noop {} {\bibfield  {journal} {\bibinfo  {journal} {Phys. Rev. E}\
  }\textbf {\bibinfo {volume} {84}},\ \bibinfo {pages} {031909} (\bibinfo
  {year} {2011})}\BibitemShut {NoStop}%
\bibitem [{\citenamefont {Lamy-Freund}\ and\ \citenamefont
  {Riske}(2003)}]{LamyFreund2003}%
  \BibitemOpen
  \bibfield  {author} {\bibinfo {author} {\bibfnamefont {M.~T.}\ \bibnamefont
  {Lamy-Freund}}\ and\ \bibinfo {author} {\bibfnamefont {K.~A.}\ \bibnamefont
  {Riske}},\ }\href@noop {} {\bibfield  {journal} {\bibinfo  {journal}
  {Chemistry and Physics of Lipids}\ }\textbf {\bibinfo {volume} {122}},\
  \bibinfo {pages} {19} (\bibinfo {year} {2003})}\BibitemShut {NoStop}%
\bibitem [{\citenamefont {Barroso}\ \emph {et~al.}(2010)\citenamefont
  {Barroso}, \citenamefont {Riske}, \citenamefont {Henriques},\ and\
  \citenamefont {Lamy}}]{Barroso2010}%
  \BibitemOpen
  \bibfield  {author} {\bibinfo {author} {\bibfnamefont {R.~P.}\ \bibnamefont
  {Barroso}}, \bibinfo {author} {\bibfnamefont {K.~A.}\ \bibnamefont {Riske}},
  \bibinfo {author} {\bibfnamefont {V.~B.}\ \bibnamefont {Henriques}}, \ and\
  \bibinfo {author} {\bibfnamefont {M.~T.}\ \bibnamefont {Lamy}},\ }\href@noop
  {} {\bibfield  {journal} {\bibinfo  {journal} {Langmuir}\ }\textbf {\bibinfo
  {volume} {26}},\ \bibinfo {pages} {13805} (\bibinfo {year}
  {2010})}\BibitemShut {NoStop}%
\bibitem [{\citenamefont {Henriques}\ \emph {et~al.}(2011)\citenamefont
  {Henriques}, \citenamefont {Germano}, \citenamefont {Lamy},\ and\
  \citenamefont {Tamashiro}}]{henriques2011}%
  \BibitemOpen
  \bibfield  {author} {\bibinfo {author} {\bibfnamefont {V.~B.}\ \bibnamefont
  {Henriques}}, \bibinfo {author} {\bibfnamefont {R.}~\bibnamefont {Germano}},
  \bibinfo {author} {\bibfnamefont {M.~T.}\ \bibnamefont {Lamy}}, \ and\
  \bibinfo {author} {\bibfnamefont {M.~N.}\ \bibnamefont {Tamashiro}},\
  }\href@noop {} {\bibfield  {journal} {\bibinfo  {journal} {Langmuir}\
  }\textbf {\bibinfo {volume} {27}},\ \bibinfo {pages} {13130} (\bibinfo {year}
  {2011})}\BibitemShut {NoStop}%
\bibitem [{\citenamefont {Carneiro}\ \emph {et~al.}(1989)\citenamefont
  {Carneiro}, \citenamefont {Henriques},\ and\ \citenamefont
  {Salinas}}]{carneiro1989}%
  \BibitemOpen
  \bibfield  {author} {\bibinfo {author} {\bibfnamefont {C.~E.~I.}\
  \bibnamefont {Carneiro}}, \bibinfo {author} {\bibfnamefont {V.~B.}\
  \bibnamefont {Henriques}}, \ and\ \bibinfo {author} {\bibfnamefont {S.~R.}\
  \bibnamefont {Salinas}},\ }\href@noop {} {\bibfield  {journal} {\bibinfo
  {journal} {Physica A: Statistical Mechanics and its Applications}\ }\textbf
  {\bibinfo {volume} {162}},\ \bibinfo {pages} {88} (\bibinfo {year}
  {1989})}\BibitemShut {NoStop}%
\bibitem [{\citenamefont {Lee}(2008)}]{lee2008}%
  \BibitemOpen
  \bibfield  {author} {\bibinfo {author} {\bibfnamefont {K.~Y.~C.}\
  \bibnamefont {Lee}},\ }\href@noop {} {\bibfield  {journal} {\bibinfo
  {journal} {Annual Review of Physical Chemistry}\ }\textbf {\bibinfo {volume}
  {59}},\ \bibinfo {pages} {771} (\bibinfo {year} {2008})}\BibitemShut
  {NoStop}%
\bibitem [{\citenamefont {Mar{\v{c}}elja}\ and\ \citenamefont
  {Wolfe}(1979)}]{marcelja1979}%
  \BibitemOpen
  \bibfield  {author} {\bibinfo {author} {\bibfnamefont {S.}~\bibnamefont
  {Mar{\v{c}}elja}}\ and\ \bibinfo {author} {\bibfnamefont {J.}~\bibnamefont
  {Wolfe}},\ }\href@noop {} {\bibfield  {journal} {\bibinfo  {journal}
  {Biochimica et Biophysica Acta (BBA) - Biomembranes}\ }\textbf {\bibinfo
  {volume} {557}},\ \bibinfo {pages} {24} (\bibinfo {year} {1979})}\BibitemShut
  {NoStop}%
\bibitem [{\citenamefont {Shida}\ and\ \citenamefont
  {Henriques}(1998)}]{shida1998}%
  \BibitemOpen
  \bibfield  {author} {\bibinfo {author} {\bibfnamefont {C.~S.}\ \bibnamefont
  {Shida}}\ and\ \bibinfo {author} {\bibfnamefont {V.~B.}\ \bibnamefont
  {Henriques}},\ }\href@noop {} {\bibfield  {journal} {\bibinfo  {journal}
  {International Journal of Modern Physics C}\ }\textbf {\bibinfo {volume}
  {09}},\ \bibinfo {pages} {801} (\bibinfo {year} {1998})}\BibitemShut
  {NoStop}%
\bibitem [{\citenamefont {Nagle}\ and\ \citenamefont
  {Tristram-Nagle}(2000)}]{nagle2000}%
  \BibitemOpen
  \bibfield  {author} {\bibinfo {author} {\bibfnamefont {J.~F.}\ \bibnamefont
  {Nagle}}\ and\ \bibinfo {author} {\bibfnamefont {S.}~\bibnamefont
  {Tristram-Nagle}},\ }\href@noop {} {\bibfield  {journal} {\bibinfo  {journal}
  {Biochimica et Biophysica Acta (BBA) - Reviews on Biomembranes}\ }\textbf
  {\bibinfo {volume} {1469}},\ \bibinfo {pages} {159} (\bibinfo {year}
  {2000})}\BibitemShut {NoStop}%
\bibitem [{\citenamefont {Callen}(1985)}]{callen}%
  \BibitemOpen
  \bibfield  {author} {\bibinfo {author} {\bibfnamefont {H.~B.}\ \bibnamefont
  {Callen}},\ }\href@noop {} {\emph {\bibinfo {title} {Thermodynamics and an
  Introduction to Thermostatistics}}},\ \bibinfo {edition} {2nd}\ ed.\
  (\bibinfo  {publisher} {Wiley},\ \bibinfo {year} {1985})\BibitemShut
  {NoStop}%
\end{thebibliography}%

\end{document}